\documentclass[10pt,prl,preprint,twocolumn]{revtex4}
\usepackage[latin1]{inputenc}
\usepackage{amsmath}
\usepackage{amsfonts}
\usepackage{amssymb}
\usepackage{makeidx}
\usepackage{graphicx}
\usepackage{tikz}
\usetikzlibrary{calc,3d}
\usepackage{hyperref}
\usepackage{array}
\usepackage{sistyle}

\hypersetup{urlcolor=blue,linkcolor=blue,citecolor=blue,colorlinks=true}

\begin{document}

\title{ Effects of  La non-magnetic impurities on the spin resonance of CeCoIn$_{5}$ }
\author{J. Panarin}
\affiliation{CEA-Grenoble, INAC-SPSMS-MDN, 17 rue des Martyrs, 38054 Grenoble Cedex 9, France}
\author{S. Raymond}
\affiliation{CEA-Grenoble, INAC-SPSMS-MDN, 17 rue des Martyrs, 38054 Grenoble Cedex 9, France}
\author{G. Lapertot}
\affiliation{CEA-Grenoble, INAC-SPSMS-MDN, 17 rue des Martyrs, 38054 Grenoble Cedex 9, France}
\author{J. M. Mignot}
\affiliation{Laboratoire Léon Brillouin, CEA/Saclay, CEA-CNRS, 91191 Gif sur Yvette, France}
\author{J. Flouquet}
\affiliation{CEA-Grenoble, INAC-SPSMS-MDN, 17 rue des Martyrs, 38054 Grenoble Cedex 9, France}

\begin{abstract}
The influence of La non magnetic impurities on the spin dynamics of CeCoIn$_{5}$ was studied by inelastic neutron scattering. In La-substituted systems, the spin resonance peak (observed at $\Omega_{res}=0.55 meV$ in the pure system) is shifted to lower energies but the ratio $\Omega_{res}/k_{B}T_{c}$ remains unchanged. The excitation broadens till it reaches 0.3 meV equal to the value of the quasi-elastic signal in the normal state. The evolution of La substitution is compared with the evolution of the magnetic resonance in Ni and Zn substituted YBa$_{2}$Cu$_{3}$O$_{7}$.
\end{abstract}

\maketitle

\begin{figure}[h]
\includegraphics[width=0.6\textwidth]{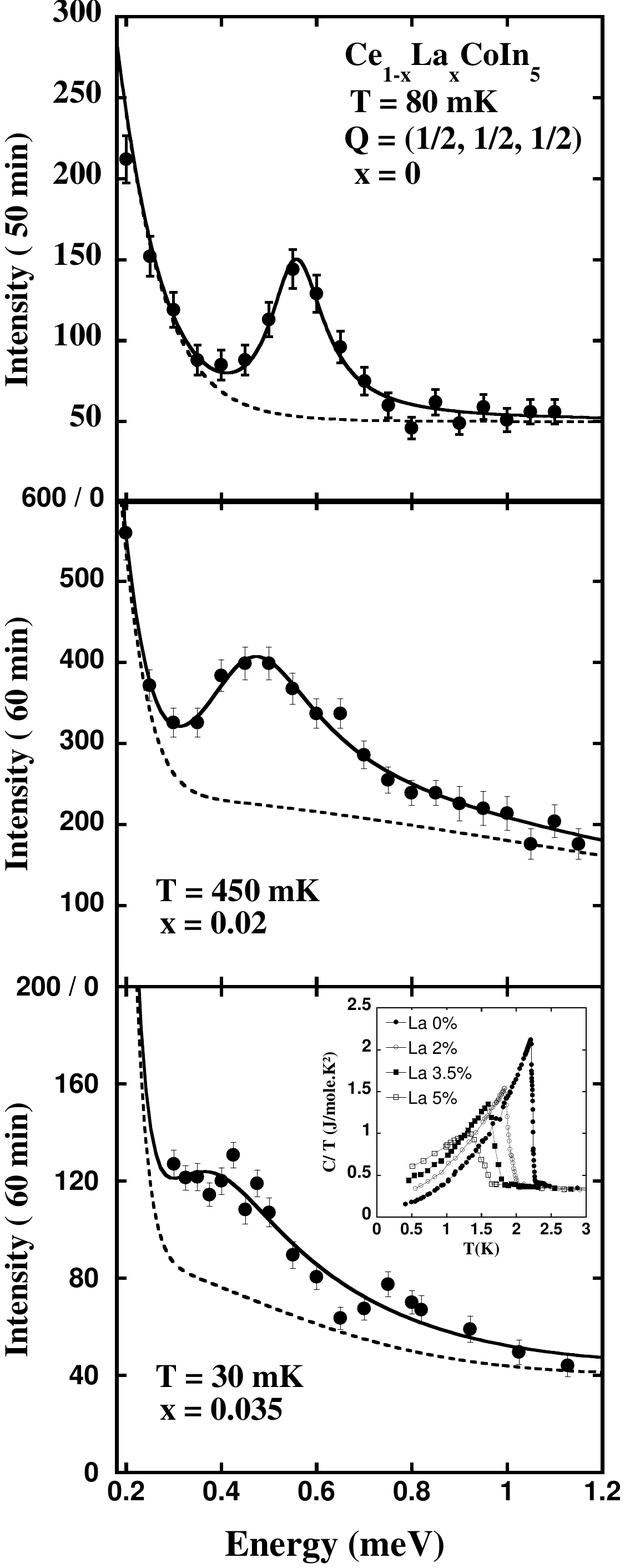}
\caption{Excitation spectra measured at $\textbf{Q} =(\frac{1}{2},\frac{1}{2},\frac{1}{2})$ for Ce$_{1-x}$La$_{x}$CoIn$_{5}$ with x = 0, 0.02 and 0.035. The solid lines are "inelastic Lorentzian" fits and the dashed line indicates the background as described in the text. The inset shows the specific heat(C$_p$) measurements of La-substitution of 0\%,2\%,3.5\% and 5\%}
\label{f1}.
\end{figure}

\begin{figure}[h]
\includegraphics[width=0.4\textwidth]{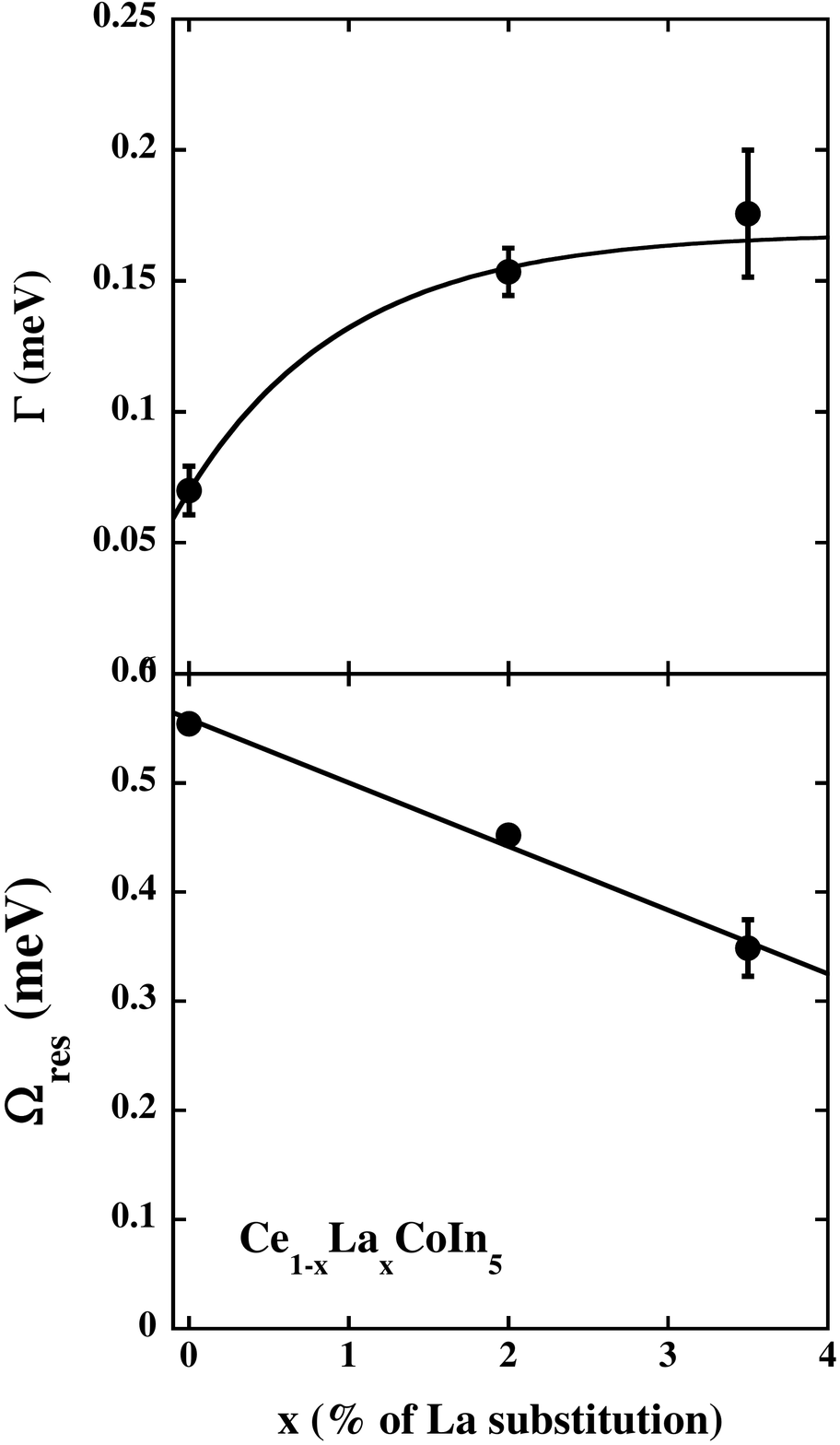}
\caption{Evolution of the relaxation rate $\Gamma$ and the resonance energy $\Omega_{res}$ as a function of the La-substitution in CeCoIn$_{5}$. The lines are a linear fit for the resonance energy and a guide for eye for the relaxation rate}
\label{f2}
\end{figure}

\begin{figure}[h]
\vspace{-0 cm}
\includegraphics[width=0.5\textwidth]{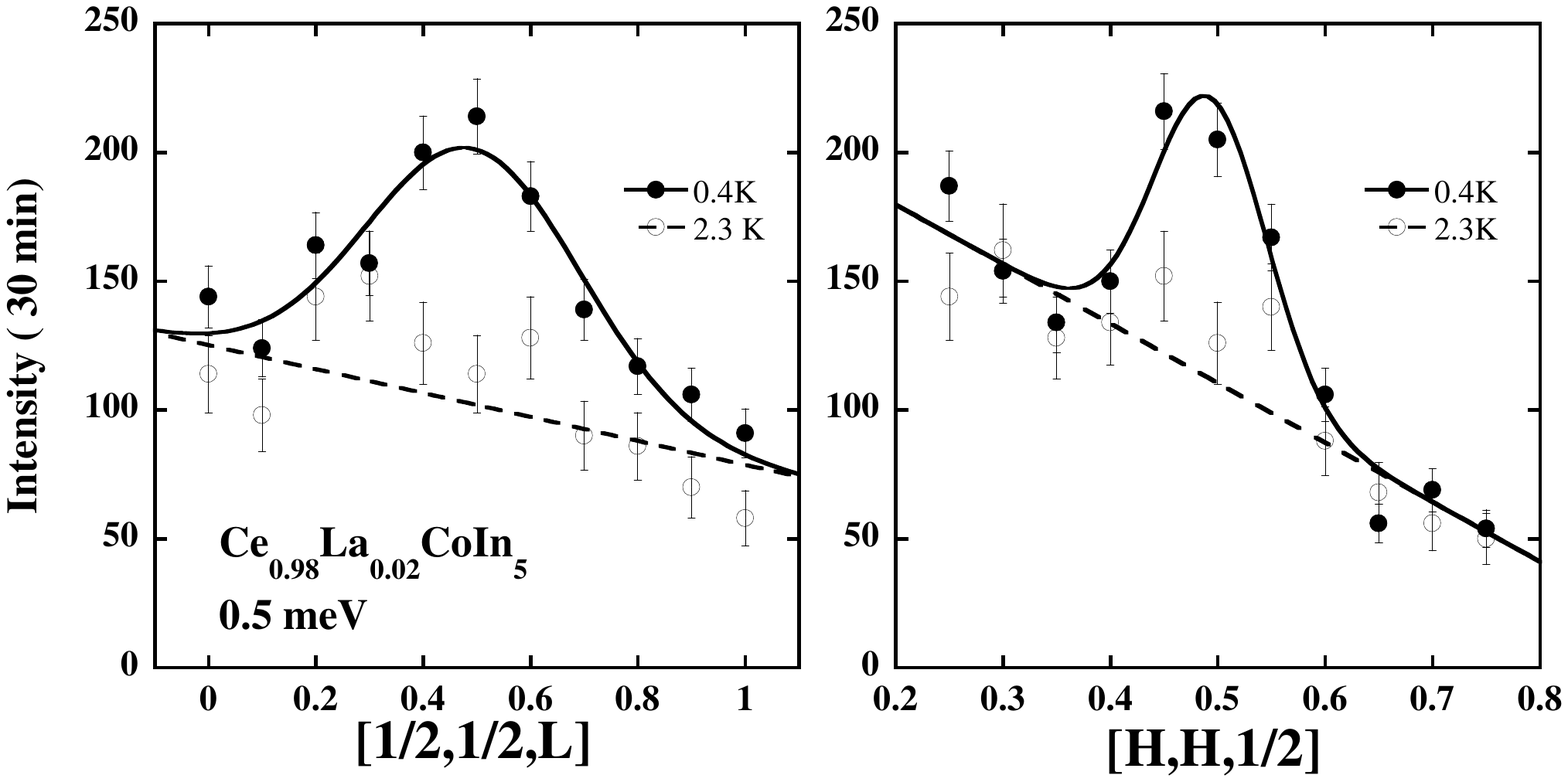}
\vspace{-8.5 cm}
\caption{Excitation spectra measured at E =0.5 meV in the direction $[\frac{1}{2},\frac{1}{2},L]$ and $[H,H,\frac{1}{2}]$  for Ce$_{0.98}$La$_{0.02}$CoIn$_{5}$. The solid line is a Gaussian fit and the dashed line indicates the background as described in the text.}
\label{f3}.
\end{figure}

Superconductivity is a macroscopic quantum state resulting from the condensation of electrons in Cooper pairs. In the case of conventionnal superconducvity the pairing mechanism is the weak electron-phonon interaction. Howewer in strongly correlated systems exhibiting a superconducting (SC) state the pairing mechanism is supposed to be of another nature. Examples of such unconventionnal superconductors are the hight temperature superconductors cuprates (HTSC), the heavy fermions compounds (HF) and the new iron-based superconductors. In these compounds, the origin of the pairing is strongly suspected to be the magnetic interaction and a seminal study of the magnetic excitation spectra by inelastic neutron scattering (INS) in the cuprate YBa$_{2}$Cu$_{3}$O$_{6+x}$  showed the appearence of a sharp excitation called magnetic spin resonance \cite{Rossat} in the SC state. Thus this was then generalized to other cuprates. Such a feedback of superconductivity on the magnetic excitation spectra was backed up by theories of a pairing mechanism of magnetic origins. The recent discovery of similar excitations in HF superconductors UPd$_{2}$Al$_{3}$\cite{Metoki} CeCoIn$_{5}$\cite{Stock} and CeCu$_{2}$Si$_{2}$\cite{Stockert} as well as the new iron superconductors \cite{Lumsden} suggests that the magnetic resonance could be a universal feature of the unconventionnal superconductors.
\\Among HF superconductors the compound CeCoIn$_5$ has the highest critical temperature, $T_{c} = 2.3 K$. It crystallizes in the tetragonal space group \textit{P4/mmm} and can be described as composed alternating CeIn$_{3}$ and CoIn layers. A quasi-2D nature is supported by de Haas van Alphen, which etablished a Fermi surface composed by nearly cylindrical sheets \cite{Shishido}. As concern the low energy magnetic excitations measured by INS, a quasielastic signal is measured above T$_c$ with a linewidth of 0.3 meV. Below T$_{c}$, the spectral shape switches from a quasielastic to a sharp inelastic peak which appears for an energy $\Omega_{res}\approx 0.55$ meV $(\approx 2.7 k_{B}T_{c})$ at the antiferromagnetic position $Q=(\frac{1}{2},\frac{1}{2},\frac{1}{2})$ \cite{Stock}.
\\The introduction of non magnetic impurities is a useful probe to investigate the microscopic nature of the SC state \cite{Balatsky}. In CeCoIn$_5$ this can be achieved by La substitution. Contrary to other types of substitution (Nd on Ce site,Cd on In cite, etc), La does not induce magnetic order but reduces only the critical temperature T$_{c}$ by $(-0.056T_{c})$/(1 \% of La substitution) \cite{Petrovic}\cite{Pham}.
\\In this Letter we report INS experiments perfomed on single crystal samples of Ce$_{1-x}$La$_{x}$CoIn$_{5}$ for x=0,0.2,0.35 and 0.5. We found that the ratio between the resonance energy $\Omega_{res}$ and T$_{c}$ remains constant when increasing the La-concentration, whereas the excitation lineshape broadens.
\\Single crystal samples were grown by the self flux method. Specific heat measurement performed using a commercial Physical Properties Measurement System (PPMS) down to 400 mK, are shown in the inset of Fig. \ref{f1}. From these measurement we deduce the T$_{c}$ of each La-substituted system studied in this Letter. Without substitution, T$_c$ is 2.3 K as previously reported \cite{Petrovic}, for x = 0.02, 0.035 and 0.05, T$_{c}$ is 1.9K, 1.7K and 1.5K respectively . The specific heat measurements are in agreement with the result obtained in Ref \cite{Tanatar}. We observe a reduction at the transition of the specific heat jump while increasing La concentration, together with a broadening of the transition and accordingly an increase of the Sommerfeld coefficient $\gamma$ for T$\rightarrow$0.
\\The INS experiments were perfomed on the cold neutron triple-axis spectrometers IN14, IN12 at Institut Laue Langevin, Grenoble and 4F2 at Laboratoire Léon Brillouin, Saclay. In the three experiments, the incident beam was provided by a vertically focusing pyrolytic graphite (PG) monochromator (double-monochromator in the case of 4F2). A liquid-nitrogen-cooled Be filter was placed before the sample in order to avoid any higher order contaminations. Measurements were perfomed with a fixed final wavevector $k_{f}= \SI{1.3}{\angstrom^{-1}}$ for IN14 and IN12, and $k_{f}=\SI{1.35}{\angstrom^{-1}}$ for 4F2. The collimations were 60'-open-open. The energy resolution determined from the full width at half-maximum (FWHM) of the incoherent signal was 0.12 meV on IN14, 0.10 on IN12 and 0.15 meV on 4F2. The different samples consisted of assemblies of about 30 single crystals of Ce$_{1-x}$La$_{x}$CoIn$_{5}$ co-aligned and glued with Fomblin oil on two thin aluminum plates. The mosaic spread of the three assemblies, as measured on a rocking curve through the (1,1,1) Bragg reflection, extend from 1 degree to 1.5 degree. The sample was put in a $^{3}$He insert for the experiments on IN14 and IN12, and in a dilution insert on 4F2 with [1,1,0] and [0,0,1] defining the scattering plane.
\\The measured neutron intensity without the background is proportionnal to the scattering funcion S(\textbf{Q},E) itself related to the imaginary part of susceptibility $\chi''(\textbf{Q},E)$.
\begin{eqnarray*}
S(\textbf{Q},E)&=&n(E,T)\;\chi''(\textbf{Q},E)
\end{eqnarray*}
\\$\chi$''was analysed using an "inelastic Lorentzian" spectral function:
\begin{eqnarray*}
\chi''(\textbf{Q},E)&=&\frac{1}{2}\left[\frac{\chi_{Q}\Gamma_{Q}E}{\left(E-\Omega_{res}\right)^2+\Gamma_{Q}^{2}}+\frac{\chi_{Q}\Gamma_{Q}E} {(E+\Omega_{res})^2+\Gamma_{Q}^{2}}\right]
\end{eqnarray*}
\\$n(E,T)= 1/(1-e^{-E/k_{B}T})$ is the detailed balance factor, $\Gamma_{Q}$ is the relaxation rate, $\Omega_{res}$ is the resonance energy and $\chi_{Q}$ is susceptbility at the wave-vector Q. All the energy spectra were taken at $\textbf{Q}=(\frac{1}{2},\frac{1}{2},\frac{1}{2})$ except the background measurements.
\\The magnetic excitation spectra measured at $Q=(\frac{1}{2},\frac{1}{2},\frac{1}{2})$ as a function of La-substitution is shown on the figure \ref{f1}. The data for x=0 are taken from Ref \cite{Panarin}. The dashed line corresponds to the background signal obtained by performing an energy scan shifted in \textbf{Q} position. For the experiment on IN12 and IN14 the background was measured at \textbf{Q} = (0.8,0.8,$\frac{1}{2}$) and for 4F2 at \textbf{Q} =(0.412,0.412,0.8). Without substitution, the resonance peak is reported in \cite{Panarin} at 0.55 meV with a relaxation rate of 0.07 meV. In 2 $\%$ La-substituted system the resonance peak shifts to $\Omega_{res}=0.45$ meV and endures a substantial broadening reaching a relaxation rate $\Gamma$ of 0.3 meV. A La-substitution of 3.5$\%$ shifts the resonance peak to $\Omega_{res}=0.35$ meV but $\Gamma$ remains constant at 0.3$\pm$0.05 meV. For this latter concentration, it is worthwhile to note that the resonance peak is no more a well-defined inelastic excitation since $\Omega_{res} \approx \Gamma$. The spectra with a 5$\%$ La-substitution (not shown here) presents no more resonance peak. Either the peak is too broad to be resolved or the excitation occurs at too low energy to be separated from the incoherent signal.
The parameters extracted from the data fit are summarized in figure \ref{f2}. In this figure the linear fit for the evolution of the resonance energy $\Omega_{res}$ as a function of La-substitution corresponds to a rate of $(-0.058\Omega_{res})$/(1\% of La substitution). As concern the relaxation rate, its increase is probably related to the diminution of the specific heat SC jump and the concomittant increase of the Sommerfeld coefficient.
\\Constant energy scans were perfomed along the \textit{a} and \textit{c}-axis around $(\frac{1}{2},\frac{1}{2},\frac{1}{2})$ in order to measure the correlation lenghts of the resonance under La-substitution. Figure \ref{f3} shows spectra measured at 0.5 meV with a subsitution of 2\%. The scans are analyzed with a gaussian lineshape. The background is determined by the measurements at hight temperature where the magnetic spectrum is no longer peaked in \textbf{Q}. The signal still peaks at $(\frac{1}{2},\frac{1}{2},\frac{1}{2})$ as for a pure compound. The correlation lenght obtained from the inverse of the gaussian half-width at half maximum are $\xi_{c}=5.1 \pm \SI{0.1}{\angstrom}$ and $\xi_{a}=11.8 \pm \SI{0.6}{\angstrom}$. In comparison with the pure compound \cite{Stock}, the correlation lengths remain similar in both directions (for a pure compound $\xi_{c}=6.5 \pm \SI{0.9}{\angstrom}$ and $\xi_{a}=9.6 \pm \SI{1}{\angstrom})$.
\\The evolution of the resonance peak of a substituted system was studied as a function of temperature. Indeed in the pure compound the resonance peak has been observed only in the SC state \cite{Stock} showing the strong coupling between the resonance excitation and superconductivity. On figure \ref{f4} we report the evolution of the $\Omega_{res}/\Omega_{res}(T=0)$ as a function of T/T$_c$ for a pure compound 0$\%$ ( extracted from the ref{\cite{Stock}) and a La-substitution of 2\%. The evolution in temperature of $\Omega_{res}$ for a 2$\%$ La-substituted system matches the evolution of a pure compound. The presence of impurities seems to have no influence on the coupling between superconductivity and resonance excitation since there is no persistance of the resonance above T$_c$.
\begin{figure}[h]
\vspace{-3 cm}
\includegraphics[width=0.4\textwidth]{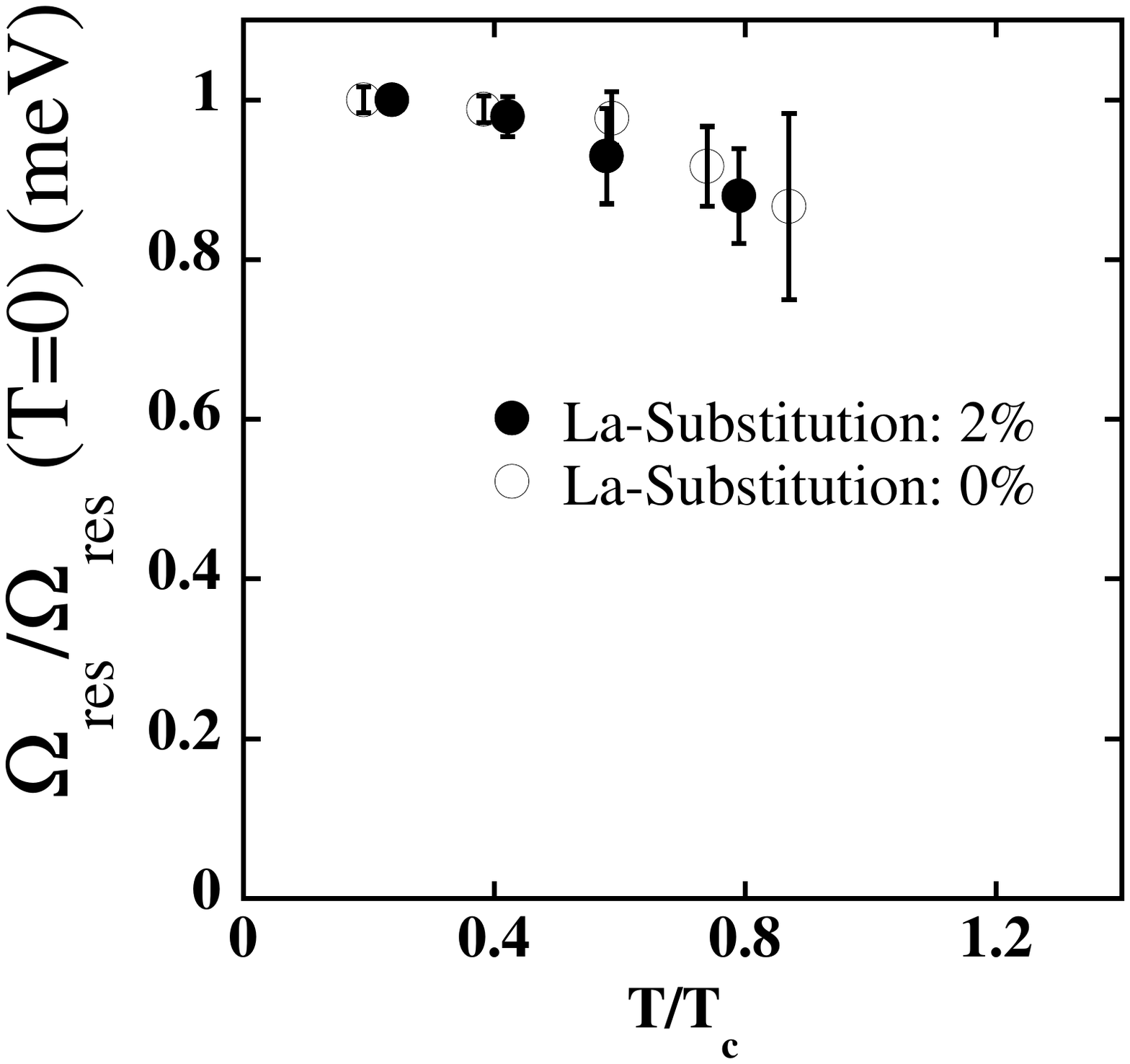}
\vspace{-2 cm}
\caption{Evolution of the resonance energy $\Omega_{res}$ of Ce$_{0.98}$La$_{0.02}$CoIn$_{5}$ and CeCoIn$_5$ from \cite{Stock} at \textbf{Q} =($\frac{1}{2}$,$\frac{1}{2}$,$\frac{1}{2}$) as function of the ratio T/T$_{c}$ with for a pure compound T$_{c}$= 2.3K and for a 2\%-substituted compound T$_{c}$= 1.9K}
\label{f4}
\end{figure}
\\When we compare the shift of $\Omega_{res}$ and the decrease of T$_{c}$ over the studied compounds, we obtain the ratio $\Omega_{res}/k_{B}T_{c} \approx 2.7$. Such a constant ratio between T$_{c}$ and $\Omega_{res}$ has already been observed in other HTSC (YBa$_{2}$Cu$_{3}$O$_{7}$ \cite{Fong}, Bi$_{2}$Sr$_{2}$CaCu$_{2}$O$_{8+x}$ \cite{He}) where the ratio is about 5.1. The linear relation in the HTSC has been established by changing the oxygen doping and so far the number of carriers but the La-substitution in CeCoIn$_{5}$ corresponds at first approximation more to the insertion of non-magnetic impurities. There are few models which have been developped concerning the effect of non-magnetic impurities on the resonance for a d-wave superconductor. To our knowledge there is only one study which describes precisely this effect,showing a decrease of the resonance energy and an increase of the resonance width \cite{Li} upon the introduction of impurities , in agreement with our results even if this theory was developped using a band structure adequate for the cuprate. 
\\Two INS experiments have been performed on the system YBCO with the substitution of Cu by Zn non-magnetic impurities and Ni magnetic impurities \cite{Fong} \cite{Sidis}.If we make an quantitative comparison with these two cases, some features drawn attention: the Zn-substitution and the Ni-substitution induce both a T$_c$ reduction but the Zn-one has a higher rate of suppression of $T_{c}: (\approx-0.13T_{c})$/(1\% of Zn substitution) in comparison of ($\approx-0.04T_{c})$/(1\% of Ni substitution) \cite{Sidis}. The influence of both impurities is totally different concerning the evolution of the spin resonance in energy and in temperature. The 1\% Zn-substitution does not shift $\Omega_{res}$ and so increases the ratio $\Omega_{res}/k_{B}T_{c}$ in comparison of a pure YBa$_{2}$Cu$_{3}$O$_{7}$ compound. Though the 3\% Ni-substitution leads to a decrease of $\Omega_{res}$ while conserving the ratio $\Omega_{res}/k_{B}T_{c}$=5.1. As concern the width of the resonance peak in energy, no noticeable increase of the relaxation rate is observed at low values of substitution \cite{Sidis} in YBa$_{2}$Cu$_{3}$O$_{7}$ with both Zn and Ni substitution contrary to our case. But the large value of the relaxation rate in YBa$_{2}$Cu$_{3}$O$_{7}$ could hide a small augmentation of the width. Moreover the magnetic signal in the 1\% Zn-substituted compound does no disappear at T$_{c}$ but half of the integrated intensity remains above T$_{c}$. As concern the Ni-case, the magnetic signal vanished upon increasing the temperature up to T$_{c}$ as the La-substituted CeCoIn$_{5}$ compound.
In conclusion the case of La-impurities in CeCoIn$_{5}$ is closer to the case of Ni magnetic impurities than Zn non-magnetic impurities in YBa$_{2}$Cu$_{3}$O$_{7}$. 
\begin{figure}[h]
\vspace{-1cm}
\includegraphics[width=0.5\textwidth]{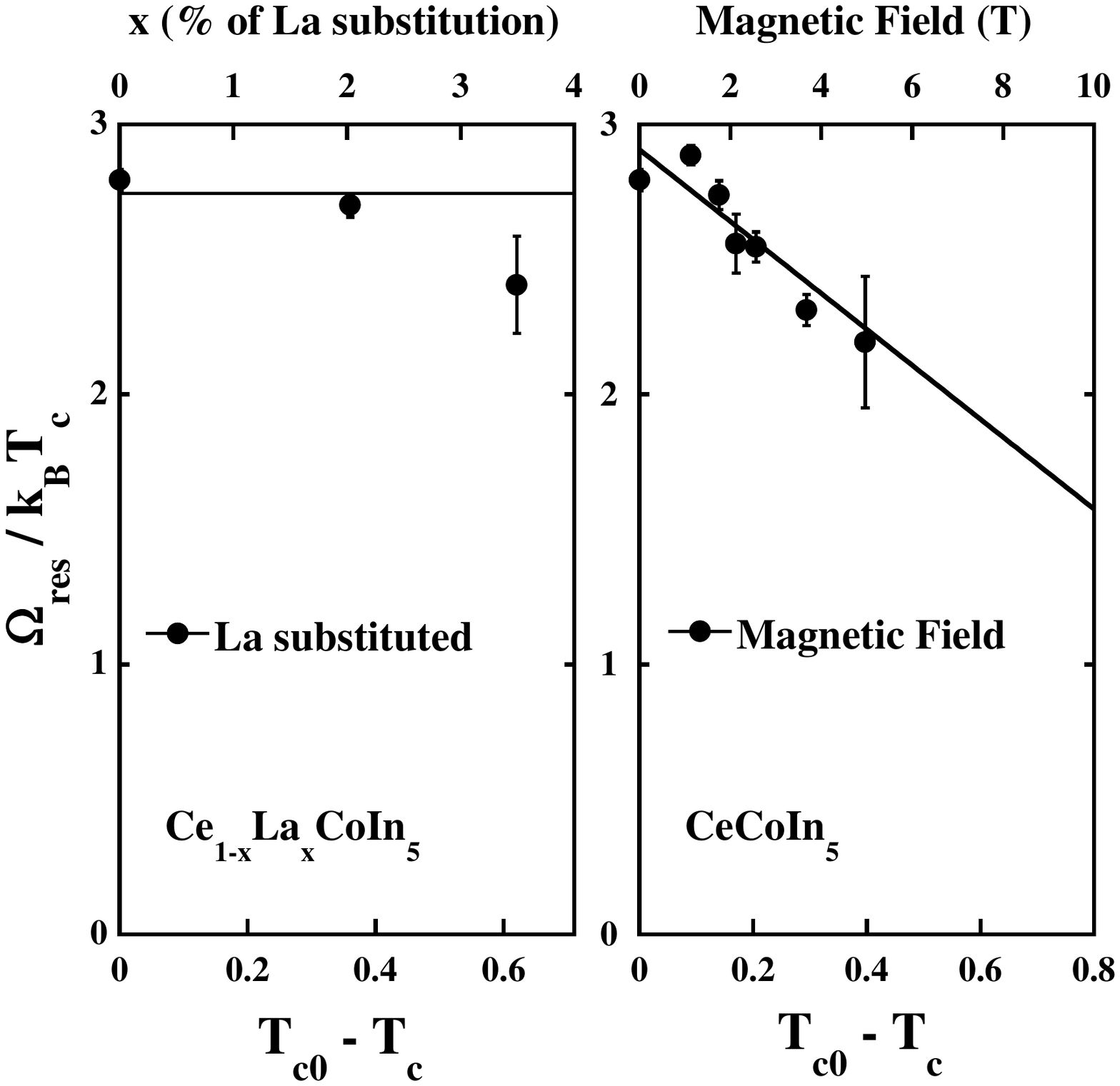}
\vspace{-4cm}
\caption{Evolution of the resonance energy $\Omega_{res}$ divided by the critical temperature as function of T$_{c0}$-T$_{c}$ controlled by La-subsitution(left panel) or magnetic field (right panel with T$_{c0}$=T$_{c}$ for pure compound without magnetic field}
\label{f5}
\end{figure}
\\Finally we compare the La-substitution with another way to tune the superconductivity, the application of a magnetic field. If we consider that the influence of the magnetic field is just a diminution of the SC gap and thus the critical temperature, the evolution of the resonance as a function of critical temperature for the La-substitution experiment (figure \ref{f5}) and for magnetic field from Ref.\cite{Panarin} shall be the same. The figure \ref{f5} proves that magnetic field has a different effect on the spin resonance that the La-substitution for which the ratio between the resonance and the critical temperature approximately remains constants with $\Omega_{res}/k_{B}T_{c} \approx 2.7$ while the application of a magnetic field diminishes this ratio. An explanation of this decrease would be to introduce a Zeeman splitting of the spin resonance under magnetic field. In this case, the lowest energy mode decreases with a linear rate while the magnetic field increases as shown in Ref. \cite{Panarin}. This behaviour is consistent with the most common models for magnetic resonance that may apply to CeCoIn$_{5}$ : the spin triplet exciton or the magnon model \cite{Chubukov}. As discussed in Ref.\cite{Panarin}, a definitive proof would be to clearly observe another mode of the multiplet at higher energy but the comparison between the La-substitution and the magnetic field effect, that span a similar range of T$_{c}$, suggests indeed a multiplet nature of the spin resonance in CeCoIn$_{5}$. 
\\The accurate study of the resonance as a function of La-substitution in CeCoIn$_{5}$ shows a almost constant ratio $\Omega_{res}/k_{B}T_{c} \approx 2.7$ with a broadening of the excitation. Theses observations are in agremment with a theoretical model developped for YBa$_{2}$Cu$_{3}$O$_{7}$ but our results are in stark contrasts with the INS experiments perfomed on YBa$_{2}$Cu$_{3}$O$_{7}$ with Zn non-magnetic impurities. Theses overall issues deserve further theoritical studies. 

\bibliographystyle{ieeetr}

\begin{thebibliography}{9}

\bibitem{Rossat} {J. Rossat-Mignod \textit{et al.}, Physica C \textbf{185}, 86-82 (1991).}

\bibitem{Metoki}{N. Metoki \textit{et al.}, J. Phys. Soc. Jpn. \textbf{66}, 2560 (1997)}

\bibitem{Stock}{C. Stock \textit{et al.}, Phys. Rev. Lett. \textbf{100}, 087001 (2008)}

\bibitem{Stockert}{O. Stockert \textit{et al.}, Physica B \textbf{403}, 973 (2008)}

\bibitem{Lumsden} {M. D. Lumsden and A. D. Christianson, J. Phys. Condens. Matter \textbf{22}, 203203(2010) and references therein}

\bibitem{Shishido} {H. Shishido  \textit{et al.}, J. Phys. Soc. Jpn. \textbf{71}, 162(2002)}

\bibitem{Balatsky} {A. V. Balatsky \textit{et al.}, Review of Modern Physics \textbf{78},373 (2006)}

\bibitem{Sidis_resonance}{Y. Sidis \textit{et al.}, C. R. Phys. \textbf{8} 745(2007)}

\bibitem{Petrovic} {C. Petrovic \textit{et al.}, J. Phys. Condens. Matter \textbf{13}, L337 (2001)}

\bibitem{Pham}{L. D. Pham \textit{et al.}, Phys. Rev. Lett. \textbf{97}, 056404 (2006)}

\bibitem{Tanatar}{M. A. Tanatar \textit{et al.}, Phys. Rev. Lett. \textbf{95}, 067002 (2005)}

\bibitem{Panarin}{J.Panarin  \textit{et al.}, J. Phys. Soc. Jpn. \textbf{78} 113706(2009)}

\bibitem{Fong}{H.F.Fong  \textit{et al.}, Phys. Rev. Lett. \textbf{82}, 1939 (1999).}

\bibitem{He}{H. F. He \textit{et al.}, Phys. Rev. Lett. \textbf{86}, 1610 (2001)}

\bibitem{Sidis}{Y.Sidis  \textit{et al.}, Phys. Rev. Lett. \textbf{84}, 5900 (2000)}

\bibitem{Li}{J. Li  \textit{et al.}, Phys. Rev. B. \textbf{58}, 2895 (1998)}

\bibitem{Chubukov} {A. V. Chubukov and L. P. Gor'kov, Phys. Rev. Lett. \textbf{101}, 147004 (2008)}


\end{thebibliography}

\end{document}